\documentclass[preprint,12pt]{elsarticle}

\usepackage{graphicx}
\usepackage[english]{babel}
\usepackage{amsmath,amssymb,amsfonts,latexsym,amsbsy}
\usepackage{gensymb}
\usepackage[colorlinks=true,linkcolor=blue,citecolor=blue,urlcolor=blue]{hyperref}
\usepackage{xcolor}
\usepackage[pagewise]{lineno}
\usepackage{siunitx}
\usepackage{url}
\usepackage{tikz}
\usepackage{adjustbox}
\usepackage{multirow}
\usepackage{threeparttable}
\usepackage{upgreek}

\journal{Classical and Quantum Gravity}

\begin{document}

\begin{frontmatter}



\title{Strategies to reduce the thermoelastic loss of multimaterial coated finite substrates}


\author[inst1]{R. Zhou}

\affiliation[inst1]{organization={Department of Materials Science and Engineering},
            addressline={University of California}, 
            city={Berkeley},
            postcode={94720}, 
            state={California},
            country={USA}}

\author[inst2]{M. Molina-Ruiz\corref{cor1}}
\ead{manelmolinaruiz@gmail.com}
\cortext[cor1]{Corresponding author}

\author[inst1,inst2]{F. Hellman}

\affiliation[inst2]{organization={Department of Physics},
            addressline={University of California}, 
            city={Berkeley},
            postcode={94720}, 
            state={California},
            country={USA}}

\begin{abstract}
Thermoelastic loss is an important energy dissipation mechanisms in resonant systems. A careful analysis of the thermoelastic loss is critical to the design of low-noise devices for high-precision applications, such as the mirrors used for gravitational-wave detectors. In this paper, we present analytical solutions to the thermoelastic loss due to thermoelasticity between different materials that are in contact. We find expressions for the thermoelastic loss of multimaterial coatings of finite substrates, and analyze its dependencies on material properties, mirror design and operating experimental conditions. Our results show that lower operating mirror temperature, thinner layers and higher number of interfaces in the coating, and the choice of the first layer of the coating that minimizes the thermal expansion mismatch with the substrate are strategies that reduce the thermoelastic loss and, therefore, diminish the thermal noise that limits the resolution in sensing applications. The results presented in this paper are relevant for the development of low-noise gravitational-wave detectors and for other experiments sensitive to energy dissipation mechanisms when different materials are in contact.
\end{abstract}


% \begin{highlights}
% \end{highlights}


\end{frontmatter}

\section{Introduction}
\label{sec:intro}
Thermal noise has been shown, through calculations derived from experimental measurements, to be one of the most significant noise sources that is limiting the sensitivity of gravitational-wave (GW) detectors~\cite{Aasi2015AdvancedLIGO, Buikema2020}. It consists of Brownian noise, which is related to mechanical loss intrinsic to the material and is the dominant contribution at room temperature~\cite{GURKOVSKY20103267}, thermorefractive noise, caused by refractive index variations due to temperature fluctuations~\cite{Evans2008Thermo-opticMeasurements}, and thermoelastic noise, caused by thermoelasticity, which is the coupling between the elastic field in the structure due to deformation and the temperature field~\cite{Braginsky1999ThermodynamicalAntennae}.

In coated substrates, as used for GW detector mirrors~\cite{Acernese2015AdvancedDetector, Aso2013InterferometerDetector}, thermoelasticity can be caused by two different mechanisms: (1) statistical temperature fluctuations, intrinsic to any material, and (2) thermal expansion mismatch between different materials in contact. Mechanism (2) will be the focus of this paper. In a vibrating structure, the periodic elastic displacement field causes a temperature gradient, where the compressed region becomes hotter and the stretched region becomes cooler. Consequently, heat transfer takes place in order to reach thermal equilibrium and yields elastic components out of phase with the input fields~\cite{Zener1937InternalReeds, Zener1938InternalFriction}. The ratio of energy dissipated due to this irreversible heat flow ($E_{diss}$) to the total energy stored in the system ($E_{stored}$) is defined as thermoelastic loss $\phi$. The loss factor, whether it is determined from a single or many dissipation mechanisms, is defined as $\phi = E_{diss} / 2\pi E_{stored}$.

The contribution to thermal noise $S_x$ from thermoelastic loss $\phi$ at a particular frequency $f$ can be found using the method proposed by Levin based on the fluctuation-dissipation theorem~\cite{PhysRevD.57.659}:
\begin{equation}\label{eq37}
    \begin{aligned}
    S_{x}(f)=\frac{2k_{B}}{\pi^2}\frac{T E_{diss}}{f F_{0}^{2}} = \frac{4k_{B}}{\pi}\frac{T E_{stored}}{f F_{0}^{2}}\phi\\
    \end{aligned}
\end{equation}
where $k_{B}$ is the Boltzmann constant, $T$ is the temperature at which $S_x(f)$ is evaluated and $F_{0}$ is the amplitude of the imposed force $F = F_{0}cos(2\pi ft)$. For further details on Eq.~\ref{eq37} and its form for a Gaussian laser beam, the specific case for GW detectors, see Refs.~\cite{PhysRevD.57.659, Bondu1998, Liu2000ThermoelasticMasses}.

Gravitational-wave detectors, such as the Laser Interferometer Gravitational-wave Observatory (LIGO)~\cite{Aasi2015AdvancedLIGO} and the Virgo interferometer~\cite{Acernese2015AdvancedDetector}, are kilometer-sized interferometers that bounce a beam of light between highly reflective optical mirrors. These mirrors at present consist of a silica substrate and a multilayer coating that alternates layers of silica and titania-doped tantala~\cite{Granata2020}. The sensitivity of such detectors improves when thermoelastic loss is reduced, thereby reducing the system thermal noise. For this reason, establishing design and experimental parameters that contribute to thermoelastic loss is important to minimizing thermal noise in GW detectors.

In 1937, Zener first published an analysis of thermoelastic damping in thin resonator beams undergoing flexural vibrations~\cite{Zener1937InternalReeds}. Lifshitz and Roukes later refined his model and solved the fundamental equations more rigorously, leading to different approximate solutions of temperature profile and hence, different expressions of the thermoelastic loss~\cite{Lifshitz2000ThermoelasticSystems}. Fejer et al. developed an independent approach for the computation of $\phi$ in mirrors for gravitational-wave detectors; they determined the energy lost due to the elastic field induced by thermal fluctuations in the coating on an infinitely thick substrate~\cite{Fejer2004ThermoelasticDetectors}. Based on the work of Fejer et al., Somiya and Yamamoto calculated the coating thermal noise considering a substrate of finite thickness and proposed a different form of solution to the thermal equations using the elastic response of a cylinder with finite thickness~\cite{Somiya2009CoatingMirror}. Fejer et al. and, separately, Somiya and Yamamoto, adopted the same approach to calculate the thermoelastic noise in mirrors with multilayered coatings: specifically, the effective medium approach (EMA), in which the multilayers are abstracted as a homogeneous medium of weighted-average physical properties~\cite{Fejer2004ThermoelasticDetectors, Somiya2009CoatingMirror}.

In this paper, and based on the model described by Fejer et al.~\cite{Fejer2004ThermoelasticDetectors}, we present analytical solutions for the thermoelastic loss of a multimaterial coated substrate of finite thickness without using the EMA. We derive expressions for $\phi$ due to thermal expansion mismatch between different layers in the coating and substrate, which allow the calculation of the associated thermoelastic noise using Eq.~\ref{eq37}. The total thermal noise due to thermoelasticity of finite size mirrors can then be estimated combining the thermoelastic noise due to mechanisms (1)~\cite{Braginsky_2003, Lovelace2018NumericallyCoatings} and (2) described above. This, however, is not the scope of this paper, and we focus on the strategies that reduce $\phi$ regarding the multilayer design of a coated finite substrate and its operating temperature. We note that $\phi$ is a ratio of energies, dimensionless and independent of the details of the laser field. The results presented in this paper show that thermoelastic loss $\phi$ can be minimized through a careful choice of variables, such as operating temperature of the mirror, thickness of layers and number of interfaces in the coating, and the choice of the first layer of the coating in contact with the substrate such as to minimize their thermal expansion mismatch.

The paper is outlined as follows: Section~\ref{math-framework} presents an overview of the key equations used to evaluate the thermoelastic loss in a substrate uniformly coated with either a single layer or a multilayer. Thermoelastic response and the consequent heat propagation normal to the interfaces are studied. We first consider a model consisting of a single layer, then expand the model to include a multilayer coating, and refine the heat equations to accommodate the alternating layers without adopting the EMA. Analytical solutions for both types of structure are derived. Finally, Section~\ref{results} presents the results calculated from the formulae of thermoelastic loss for various cases and discusses how material properties and external factors impact the thermoelastic loss. Particular attention is paid to the multilayer structure, where the structure of the coating (layers thickness and number of interfaces) is preserved.

\section{Mathematical framework for the thermoelastic loss}\label{math-framework}
\begin{figure}[!h]
    \centering
    \includegraphics[width=8.0cm]{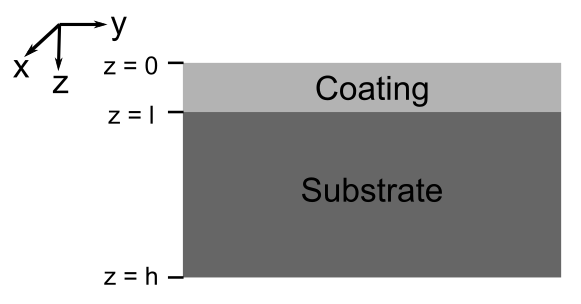}
    \caption{Illustration of the homogeneous system studied by our model: a substrate coated with a film.}
    \label{figure-singlelayermodel}
\end{figure}

To derive an expression for the thermoelastic loss in a coated substrate, we consider a film of thickness $l$ on a substrate with a thickness of $h-l$, as illustrated in Fig.~\ref{figure-singlelayermodel}. We define the surface normal to be in the $z$-direction. The film's surface is located at $z=0$ and its interface with the substrate at $z=l$, the substrate extends from $z=l$ to $z=h$, where $h \gg l$. Our model assumes that both the film and the substrate are homogeneous, i.e., that their physical properties do not change throughout their volume, that there is no temperature variation in the $x$-$y$ plane, and that the transverse dimensions are much larger than the longitudinal ones (thicknesses of film and substrate), therefore, only the thermal diffusion along $z$-direction needs to be taken into account.

The energy loss due to thermoelasticity is calculated for two states of stress, in-plane stress and normal stress. Detailed derivations are presented in the following sections. Thermoelastic loss is a measure of energy dissipated due to the coupling between the deformation and thermal fields. Hence first, the relationship between the intrinsic elastic field and the temperature field produced by it needs to be determined. By defining the boundary conditions and solving the equilibrium equations, the thermal strain and the associated dissipated power can be found. Subsequently, the elastic energy stored in the coating can be obtained from the applied stress through the use of theory of elasticity. Finally, based on the definition of thermoelastic loss $\phi$, an expression of $\phi$ for the system can be found.

\subsection{Formulation of the thermal field}\label{formulationthermalfield}
The dynamic deformation of a body causes temperature variations. We use the linear heat equation along the $z$-direction to find the temperature distribution coupled to the input elastic field~\cite{L.D.Landau1986TheoryElasticity},
\begin{equation}\label{eq1}
    \frac{\partial \theta_j}{\partial t}-\kappa_j\frac{\partial^2\theta_j}{\partial z^2}=-\frac{E_j\alpha_{j}T}{(1-2\nu_j)C_j}\frac{\partial}{\partial t}\sum\epsilon_{0,j}
\end{equation}
where $\theta_j(t,z)$ is the time/position-varying temperature, $\kappa_j$ is the thermal diffusivity, $E_j$ is the Young's modulus, $\alpha_{j}$ is the coefficient of linear thermal expansion, $T$ is the background temperature, $\nu_j$ is the Poisson's ratio, $C_j$ is the specific heat capacity per unit volume and $j = f$, $s$ indicates quantities evaluated in the film ($f$) and in the substrate ($s$), respectively. GW mirrors are exposed to high-intensity laser beams; therefore, the mirror thermal field depends on both elastic field and optical absorption. In the model presented in this paper, we only study the dependence of the thermal field with the elastic field. The work on photothermal transfer function by Ballmer~\cite{Ballmer2015PhotothermalMeasurements} serves as a basis for calculating the heat flow and thermal expansion caused by optical absorption.

Taking $\theta_j(t,z)$ to be in the form of $\theta_{j}(z)\exp(i\omega t)$ and the strain as $\epsilon_0\exp(i\omega t)$, where $\omega = 2\pi f$ is the angular frequency, Equation~\ref{eq1} becomes,
\begin{equation}\label{eq2}
    i\omega\theta_j-\kappa_j\frac{\partial^2\theta_j}{\partial z^2}=-i\omega\beta_j
\end{equation}
where $\beta_j = (E_j \alpha_{j} T) / [(1-2\nu_j) C_j] \sum\epsilon_{0,j}$.
The boundary conditions for heat fluxes are defined as follows,
\begin{equation}\label{eq3}
    \begin{aligned}
    \frac{\partial\theta_f}{\partial z}\Bigr|_{\substack{z=0}}&=0 \text{ ,} \\
    k_f\frac{\partial\theta_f}{\partial z}\Bigr|_{\substack{z=l}}&=k_s\frac{\partial\theta_s}{\partial z}\Bigr|_{\substack{z=l}} \text{ ,} \\
    \frac{\partial\theta_s}{\partial z}\Bigr|_{\substack{z=h}}&=0 \\
    \end{aligned}
\end{equation}
where $k$ represents the thermal conductivity.

As previously proposed by Fejer et al.~\cite{Fejer2004ThermoelasticDetectors}, the final solution to the heat equation consists of two parts, a particular ($p$) solution of thermal field, $\theta_{p,j}(z)$, that is coupled to the strain and satisfies Eq.~\ref{eq2}, and a specific ($s$) solution, $\theta_{s,j}(z)$, that meets the boundary conditions. It can be written as
\begin{equation}\label{eq4}
    \theta_j = \theta_{p,j}(z) + \theta_{s,j}(z)
\end{equation}
We solve for the thermal fields consistent with the boundary conditions (Eq.~\ref{eq3}); the calculation is shown in~\ref{sec:appendix-thermalfields}. The thermal fields in the film and the substrate along the $z$ direction are
\begin{equation}\label{eq5}
    \begin{aligned}
    \theta_{f} &= -\beta_{f}+\frac{\Delta\beta\times cosh(\gamma_{f}z)}{\cosh(\gamma_{f}l) + R\sinh(\gamma_{f}l)\coth(q)} \text{ ,} \\
    \theta_{s} &= -\beta_{s}-\frac{\Delta\beta R\times cosh[\gamma_{s}(h-z)]}{\coth(\gamma_{f}l)\sinh(q) + R\cosh(q)}
    \end{aligned}
\end{equation}
where $q = \gamma_{s}(h-l)$, $\Delta\beta = \beta_f-\beta_s$, $\gamma_j=(1+i)\sqrt{\pi f C_{j}/k_{j}}$ and $R=(k_{f}\gamma_{f})/(k_{s}\gamma_{s})=\sqrt{(k_{f}C_{f})/(k_{s}C_{s})}$.

To determine the induced stress field $\epsilon_{1}$ and strain field $\sigma_{1}$ by oscillatory thermal fluctuation, we assume that any expansion/contraction along the $z$-direction is not restricted, and therefore, $\epsilon_{1,zz}=\alpha_{j}\theta_{j}$, $\sigma_{1,zz}=0$. When considering the in-plane thermal expansion, one assumption made in the calculation is that the substrate expands freely, effectively uninfluenced by the film; however, the in-plane expansion of the film is constrained. Moreover, the film is assumed to have a uniform strain of $\alpha_s\theta_s(z=l)$, which is the expansion of the substrate at the interface with the film. This approximation is reasonable if the substrate is much thicker than the film, i.e., $(h-l) \gg l$. Since the free expansion in plane of the film is forbidden, stress is developed in the interior of the film and the stress level depends on the mismatch between the thermal expansion of the film and the substrate. Taking these into consideration, we find the elastic fields due to the thermal mismatch as
\begin{equation}\label{eq6}
    \begin{aligned}
    \epsilon_{1,ii,j} &= A_{1,ii,j}\alpha_j\theta_j \text{ ,} \\
    \sigma_{1,ii,j} &= B_{1,ii,j}\alpha_j\theta_j
    \end{aligned}
\end{equation}
The matrices $A_{1,ii,j}$ and $B_{1,ii,j}$ are defined as
\begin{equation}\label{eq7}
    \begin{aligned}
    A_{1,ii,f} &= \begin{pmatrix} \alpha_{s}\theta_{s,l} & 0 & 0\\ 0 & \alpha_{s}\theta_{s,l} & 0\\ 0 & 0 & \alpha_{f}\theta_{f} \end{pmatrix} \text{ ,} \\
    B_{1,ii,f} &= \begin{pmatrix} -b & 0 & 0\\ 0 & -b & 0\\ 0 & 0 & 0 \end{pmatrix} \text{ ,} \\
    A_{1,ii,s} &= \begin{pmatrix} \alpha_{s}\theta_{s} & 0 & 0\\ 0 & \alpha_{s}\theta_{s} & 0\\ 0 & 0 & \alpha_{s}\theta_{s} \end{pmatrix} \text{ ,} \\
    B_{1,ii,s} &= \begin{pmatrix} 0 & 0 & 0\\ 0 & 0 & 0\\ 0 & 0 & 0 \end{pmatrix}
    \end{aligned}
\end{equation}
where $b = E_{f}(\alpha_{f}\theta_{f}-\alpha_{s}\theta_{s,l})$, and $\theta_{s,l}$ represents $\theta_s(z=l)$.

\subsection{Applied elastic field and elastic energy}\label{appliedelasticfield}
In order to calculate the oscillatory thermal field caused by an oscillatory stress or a vibration at a frequency $\omega$, we need to first determine the stress and strain states. Here two potential cases are considered: 1) in-plane stress and 2) normal stress to the surface. The latter one is relevant to mechanical loss measurements and LIGO operation. \textcolor{black}{Solutions to other stress states can be attained by a sum of the solutions based on these two stress fields.} Using the elastic boundary conditions, we find the driving elastic fields in the film, denoted by the subscript $0$.

\subsubsection{Stress parallel to the coated surface}
The boundary conditions for stress parallel to the coated surface are: $\sigma_{0,xx} = \sigma_{0,yy} = \sigma_{0,\parallel}$ (with the assumption of symmetry), and $\sigma_{0,zz} = 0$. Using Hooke's law, we find
\begin{equation}\label{eq8}
    \begin{aligned}
    \epsilon_{0,xx,j}=&\epsilon_{0,yy,j}=\frac{1-\nu_j}{E_j}\sigma_{0,\parallel} \text{ ,} \\
    \epsilon_{0,zz,j}&=\frac{-2\nu_j}{E_j}\sigma_{0,\parallel}
    \end{aligned}
\end{equation}

Summarizing the results in matrix form, the stress and strain fields can be expressed in the forms of
\begin{equation}\label{eq9}
    \begin{aligned}
    \epsilon_{0,ii,\parallel,j}&=a_{0,ii,j}\sigma_{0,\parallel} \text{ ,} \\
    \sigma_{0,ii,\parallel,j}&=b_{0,ii,j}\sigma_{0,\parallel}
    \end{aligned}
\end{equation}
where
\begin{equation}\label{eq10}
    \begin{aligned}
    a_{0,j} &= \begin{pmatrix} \frac{1-\nu_j}{E_j} & 0 & 0\\ 0 & \frac{1-\nu_j}{E_j} & 0\\ 0 & 0 & \frac{-2\nu_j}{E_j} \end{pmatrix} \text{ ,} \\
    b_{0,j} &= \begin{pmatrix} 1 & 0 & 0\\ 0 & 1 & 0\\ 0 & 0 & 0 \end{pmatrix}
    \end{aligned}
\end{equation}
Therefore, the elastic energy stored per unit area is given by
\begin{equation}\label{eq11}
    \begin{aligned}
    E_{stored,\parallel,j} &= \frac{1}{2}\sigma\epsilon\times l\\
    &= \frac{1}{2}\sigma_{0,\parallel,j}^{2}l\sum b_{0,j}a_{0,j}\\
    &= \frac{1}{2}\sigma_{0,\parallel,j}^{2}l\times\frac{2(1-\nu_j)}{E_j} 
    \end{aligned}
\end{equation}

\subsubsection{Stress perpendicular to the coated surface}
For stress perpendicular to the coated surface, the boundary conditions are defined as: $\sigma_{0,zz} = \sigma_{0,\perp}$, $\sigma_{0,xx}, \sigma_{0,yy} \neq 0$; $\epsilon_{0,zz} \neq 0$, $\epsilon_{0,xx} = \epsilon_{0,yy} = 0$. With the Hooke's law expression, we obtain
\begin{equation}\label{eq12}
    \begin{aligned}
    \sigma_{0,xx,j}&=\sigma_{0,yy,j}=\frac{\nu_j}{1-\nu_j}\sigma_{0,\perp} \text{ ,} \\
    \epsilon_{0,zz,j}&=\frac{\sigma_{0,\perp}}{E_j}\frac{(1-2\nu_j)(1+\nu_j)}{1-\nu_j}
    \end{aligned}
\end{equation}
The collective results are,
\begin{equation}\label{eq13}
    \begin{aligned}
    \epsilon_{0,ii,\perp,j}&=c_{0,ii,j}\sigma_{0,\perp} \text{ ,} \\
    \sigma_{0,ii,\perp,j}&=d_{0,ii,j}\sigma_{0,\perp}
    \end{aligned}
\end{equation}
where
\begin{equation}\label{eq14}
    \begin{aligned}
    c_{0,j} &= \begin{pmatrix} 0 & 0 & 0\\ 0 & 0 & 0\\ 0 & 0 & \frac{(1-2\nu_j)(1+\nu_j)}{1-\nu_j}\frac{1}{E_j} \end{pmatrix} \text{ ,} \\
    d_{0,j} &= \begin{pmatrix} \frac{\nu_j}{1-\nu_j} & 0 & 0\\ 0 & \frac{\nu_j}{1-\nu_j} & 0\\ 0 & 0 & 1 \end{pmatrix}
    \end{aligned}
\end{equation}
Hence, the amount of elastic energy stored per unit area is given by
\begin{equation}\label{eq15}
\begin{aligned}
    E_{stored,\perp,j} &= \frac{1}{2}\sigma\epsilon\times l\\
    &= \frac{1}{2}\sigma_{0,\perp,j}^{2}l\sum d_{0,j}c_{0,j}\\
    &= \frac{1}{2}\sigma_{0,\perp,j}^{2}l\times\frac{(1-2\nu_j)(1+\nu_j)}{E_j(1-\nu_j)} 
\end{aligned}
\end{equation}
Additionally, specific elastic fields can be substituted into this model by defining $\epsilon_{0,ii,j}$ and $\sigma_{0,ii,j}$ to obtain $E_{stored,j}$ for these specific cases.

\subsection{Energy dissipation and thermoelastic loss}\label{energydissipation}
We first consider the energy dissipated within the film. The rate of energy dissipation per unit volume in a deformed body, in this case, the film, is defined as
\begin{equation}\label{eq16}
    p_{diss,f} = \frac{\sigma}{2}\frac{\partial\epsilon}{\partial t}
\end{equation}
which has to be a real function. The overall film stress and strain in the case of plane stress are described as
\begin{equation}\label{eq17}
    \begin{aligned}
    \epsilon_{f,ii,\parallel} &= a_{0,ii}\epsilon_{0,\parallel} + A_{1,ii,f}\alpha_f\theta_f \text{ ,} \\
    \sigma_{f,ii,\parallel} &= b_{0,ii}\sigma_{0,\parallel} + B_{1,ii,f}\alpha_f\theta_f
    \end{aligned}
\end{equation}
As for the case of stress perpendicular to the coated surface, we write
\begin{equation}\label{eq18}
    \begin{aligned}
    \epsilon_{f,ii,\perp} &= c_{0,ii}\epsilon_{0,\perp} + A_{1,ii,f}\alpha_f\theta_f \text{ ,} \\
    \sigma_{f,ii,\perp} &= d_{0,ii}\sigma_{0,\perp} + B_{1,ii,f}\alpha_f\theta_f
    \end{aligned}
\end{equation}
The oscillatory stress and strain fields have a time dependence
\begin{equation}\label{eq19}
    \begin{aligned}
    \epsilon = \epsilon_{f,ii}\times e^{i\omega t} \text{ ,} \\
    \sigma = \sigma_{f,ii}\times e^{i\omega t}
    \end{aligned}
\end{equation}
Hence, $p_{diss,f}$ can be written as
\begin{equation}\label{eq20}
    \begin{aligned}
    p_{diss,f}&=\sum\frac{1}{2}\frac{\partial\sigma_{f,ii}\epsilon_{f,ii}}{\partial t}\\
    &=\omega\sum \operatorname{Im}[\sigma_{f,ii}\epsilon_{f,ii}]\\
    &=\omega\sum \operatorname{Im}[\sigma_{0,ii}\epsilon_{0,ii} + \sigma_{1,ii}\epsilon_{0,ii} \\
    &\quad + \sigma_{0,ii}\epsilon_{1,ii}+\sigma_{1,ii}\epsilon_{1,ii}]\\
    &=\omega\sum \operatorname{Im}[\sigma_{1,ii}\epsilon_{0,ii}+\sigma_{0,ii}\epsilon_{1,ii}]  
    \end{aligned}
\end{equation}
since $\sigma_{0,ii}\epsilon_{0,ii}$ is real, and $\sigma_{1,ii}\epsilon_{1,ii}$ can be neglected because the induced elastic fields are significantly smaller than the input fields.
Therefore, the energy dissipated per unit area is
\begin{equation}\label{eq21}
E_{diss,f}=\frac{2\pi}{\omega}\int_{0}^{l}p_{diss,f}dz
\end{equation}
Substituting Eq.~\ref{eq20} and the expressions of thermal and applied elastic fields, previously derived in Sections~\ref{formulationthermalfield} and \ref{appliedelasticfield}, into Eq.~\ref{eq21}, we obtain the total energy dissipated in the film and the substrate for parallel and perpendicular stress fields, respectively:
\begin{equation}\label{eq22}
    \begin{aligned}
    E_{diss,total,\parallel} &= 2\pi\sigma_{0}\times \operatorname{Im}\left[(2\nu_{f}-2)\alpha_{f}\theta_{1,f}\gamma_{f}^{-1}\sinh(\gamma_f l) \right. \\ &\quad +(4-2\nu_{f})\alpha_{s}\theta_{1,s}\cosh(\gamma_{s}(h-l))\times l \\
    &\quad -2\alpha_{s}\theta_{1,s}\gamma_{s}^{-1}\sinh(\gamma_{s}(h-l))\Big. \Big]
    \end{aligned}
\end{equation}
\begin{equation}\label{eq23}
    \begin{aligned}
    E_{diss,total,\perp} &= 2\pi\sigma_{0}\times \operatorname{Im}\left[\alpha_{f}\theta_{1,f}\gamma_{f}^{-1}\sinh(\gamma_f l) \right.\\
    &\quad +\frac{2\nu_{f}}{1-\nu_{f}}\alpha_{s}\theta_{1,s}\cosh[\gamma_{s}(h-l)]\times l\\
    &\quad -\frac{1+\nu_{s}}{1-\nu_{s}}\alpha_{s}\theta_{1,s}\gamma_{s}^{-1}\sinh[\gamma_{s}(h-l)]\Big. \Big]
    \end{aligned}
\end{equation}
The detailed derivation is included in~\ref{sec:appendix-energydissipation}. 

The loss factor is generally defined as the ratio of energy dissipated per radian to the potential energy in a cycle. In this case, the potential energy is the elastic energy stored in the strained material. The consideration of an infinite substrate prevents the analysis of the elastic energy stored in the substrate, and only the strain energy of the coating can be considered. Therefore, to allow a direct addition/subtraction of all loss components (mechanical and thermoelastic), the denominator has to be consistently defined for these loss factors as the elastic energy of the system (coating and substrate). In this paper, we consider the stored potential energy in both the coating and the substrate.

According to Eqs.~\ref{eq11} and~\ref{eq15}, the total stored energy in the system can be written in the form of
\begin{equation}\label{eq24}
    \begin{aligned}
    E_{stored,total,\parallel}&=\frac{1}{2}\sigma_0^{2}l\times\frac{2(1-\nu_f)}{E_{f}} \\
    &\quad+ \frac{1}{2}\sigma_0^{2}(h-l)\times\frac{2(1-\nu_{s})}{E_{s}} \text{ ,} \\
    E_{stored,total,\perp}&=\frac{1}{2}\sigma_0^{2}l\times\frac{(1-2\nu_f)(1+\nu_f)}{E_f(1-\nu_f)} \\
    &\quad+ \frac{1}{2}\sigma_0^{2}(h-l)\times\frac{(1-2\nu_s)(1+\nu_s)}{E_s(1-\nu_s)}
    \end{aligned}
\end{equation}
Therefore, the thermoelastic loss $\phi$ can thus be calculated from
\begin{equation}\label{eq25}
    \begin{aligned}
    \phi_{\parallel} &= \frac{\left|E_{diss,total,\parallel}\right|}{2\pi\times E_{stored,total,\parallel}} \text{ ,} \\
    \phi_{\perp} &= \frac{\left|E_{diss,total,\perp}\right|}{2\pi\times E_{stored,total,\perp}}
    \end{aligned}
\end{equation}

Typically the substrate is significantly thicker than the film, i.e., $(h-l) \gg l$, implying that the total stored energy is mostly contained in the substrate (the second terms of Eq.~\ref{eq24}). As a result, by using a thicker substrate, which leads to a larger elastic energy stored in the substrate (while having little impact on the temperature gradient and inducing only a small increase in the energy lost per cycle due to thermoelasticity), we see a decrease in $\phi$. It should be noted that this is different from the dilution factor~\cite{Li2014} considered in mechanical loss measurements from the GeNS method~\cite{Cesarini2009}. In the GeNS method, the mechanical loss of a coating is determined as the difference between the total (substrate + coating) and the substrate mechanical losses weighted by the dilution factor, which is defined as the ratio between the energy stored in the coating and the total energy of the system~\cite{Li2014}. Since the thermoelastic loss that we report in this paper is caused by the thermal expansion mismatch between different materials, it is due to interactions between layers, i.e., $\phi$ cannot be determined as the combined contribution from single layers. Somiya and Yamamoto showed that $\phi$ decreases with the substrate thickness and hits a plateau at a certain thickness~\cite{Somiya2009CoatingMirror}. As per the definition of thermoelastic loss (Eq.~\ref{eq25}), the thicker the substrate the larger the system stored energy, while the energy dissipated remains almost unchanged. Thus, using a thicker substrate would lead to a lower thermoelastic loss caused by thermal expansion mismatch between different materials. The thermoelastic loss plateau discussed above is a direct consequence of only taking into account the film elastic energy. The substrate effect will be discussed in more detail in Section~\ref{results}.

\subsection{Multilayer coating on a finite substrate}\label{multilayer}
\begin{figure}[!h]
    \centering
    \includegraphics[width=8.0cm]{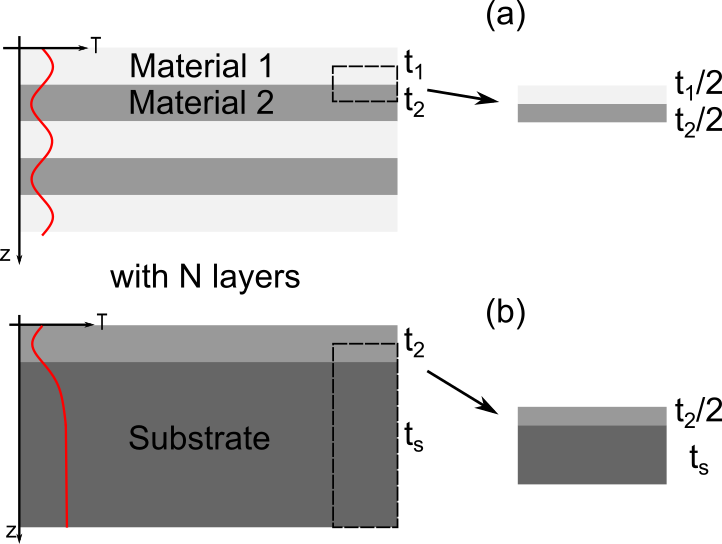}
    \caption{Illustration of a substrate coated with a multilayer showing the model used to calculate (a) $\phi$ between layers, and (b) $\phi$ between the first deposited layer and the substrate, where $N$ is the number of layers in the coating, $t_{1}$ the thickness of a single layer made of Material 1, $t_{2}$ the thickness of a single layer made of Material 2, and $t_{s}$ the substrate thickness. The red lines show an example of the temperature profile across coating and substrate.}
    \label{multilayer_model}
\end{figure}

In this section, we consider a coating made of alternating layers of two materials attached to a homogeneous substrate, which resembles the dielectric mirror coating used in gravitational-wave detectors, as depicted in Fig.~\ref{multilayer_model}. We make the assumptions that i) the outer surface is not subjected to any heat flux, namely, that heat transfer only takes place within the films and substrate, and ii) each component is only affected by its nearest neighbors. These assumptions are likely to be valid in most cases as the temperature fluctuation caused by the driving stress is reasonably small so that there would not exist a large temperature difference across the coated body. Therefore, we study the thermoelastic dissipation dividing the system in two regions: along 1) the coating layers, and 2) the first deposited layer and the substrate.

To find the energy dissipation between coating layers made of two different materials, we begin by considering the thermal field in a single layer and assuming that its derivative is zero at the center, a direct consequence of continuity. We assume that each layer is homogeneous and the system is in steady state. The thermal fields are then solved using the boundary conditions and a model consisting of two half-layers [depicted in Fig.~\ref{multilayer_model}(a)], where the temperature profile and heat flux are deemed symmetric.

The thermal field solutions \textcolor{black}{for the stack illustrated in Fig.~\ref{multilayer_model}(a)} are similar to those derived in Section~\ref{formulationthermalfield}, with the boundary conditions now being
\begin{equation}\label{eq26}
    \begin{aligned}
    \frac{\partial\theta_{f_{1}}}{\partial z}\Bigr|_{\substack{z=\frac{t_{1}}{2}}}&=0 \text{ ,} \\
    k_{f_{1}}\frac{\partial\theta_{f_{1}}}{\partial z}\Bigr|_{\substack{z=t_{1}}} &=k_{f_{2}}\frac{\partial\theta_{f_{2}}}{\partial z}\Bigr|_{\substack{z=t_{1}}} \text{ ,} \\
    \frac{\partial\theta_{f_{2}}}{\partial z}\Bigr|_{\substack{z=t_{1} + \frac{t_{2}}{2}}}&=0\\
    \end{aligned}
\end{equation}
\textcolor{black}{where $z=t_{1}/2$ is the center of a layer of Material 1, $z=t_{1}$ is the interface between Material 1 and Material 2, $z=t_{1}+t_{2}/2$ is the center of a layer of Material 2. The thermal profile derived for this particular stack is repeated due to periodicity in the multilayer coating.}

We solve for the thermal fields in the two materials and obtain
\begin{equation}\label{eq27}
    \begin{aligned}
    \theta_{f_{1}} &= \theta_{p,f_{1}}+\theta_{1,f_{1}}\cosh(\gamma_{f_{1}}z) \text{ ,} \\
    \theta_{f_{2}} &= \theta_{p,f_{2}}+\theta_{1,f_{2}}\cosh \left[ \gamma_{f_{2}} \left( \frac{t_{1}}{2}+\frac{t_{2}}{2}-z\right) \right]
    \end{aligned}
\end{equation}
where $\theta_{1,f_{1}}$ and $\theta_{1,f_{2}}$ are defined as
\begin{equation}\label{eq28}
    \begin{aligned}
    \theta_{1,f_{1}} &= \frac{\Delta\beta^{'}}{\cosh\left(\frac{\gamma_{f_{1}}t_{1}}{2}\right)+R^{'}\sinh\left(\frac{\gamma_{f_{1}}t_{1}}{2}\right)\coth\left(\frac{\gamma_{f_{2}}t_{2}}{2}\right)} \text{ ,} \\
    \theta_{1,f_{2}} &=
    -\frac{\Delta\beta^{'} R^{'}}{\coth\left(\frac{\gamma_{f_{1}}t_{1}}{2}\right)\sinh\left(\frac{\gamma_{f_{2}}t_{2}}{2}\right)+R^{'}\cosh\left(\frac{\gamma_{f_{2}}t_{2}}{2}\right)}
    \end{aligned}
\end{equation}
where $\Delta\beta^{'} = \beta_{f_{1}}-\beta_{f_{2}}$ and $R^{'}=(k_{f_{1}}\gamma_{f_{1}})/(k_{f_{2}}\gamma_{f_{2}})=\sqrt{(k_{f_{1}}C_{f_{1}})/(k_{f_{2}}C_{f_{2}})}$.
The energy dissipated for parallel and perpendicular fields in the adjacent films (illustrated in Fig.~\ref{multilayer_model}(a), indicated by subscript $a$, respectively, is thus written as
\begin{equation}\label{eq29}
    \begin{aligned}
    E_{diss,a,\parallel} &= 2\pi\sigma_{0}\times \operatorname{Im}\left[(2\nu_{f_1}-2)\frac{\alpha_{f_1}\theta_{1,f_1}}{\gamma_{f_1}}\sinh\left( \frac{\gamma_{f_1}t_{1}}{2}\right) \right.\\
    &\quad +(4-2\nu_{f_1})\alpha_{s}\theta_{1,s}\cosh(\gamma_{s}t_{s})\times\frac{t_{1}}{2}\\
    &\quad -(2-2\nu_{f_2})\frac{\alpha_{f_2}\theta_{1,f_2}}{\gamma_{f_2}}\sinh\left( \frac{\gamma_{f_2}t_{2}}{2}\right) \\
    &\quad +(4-2\nu_{f2})\alpha_{s}\theta_{1,s}\cosh(\gamma_{s}t_{s})\times\frac{t_{2}}{2}\bigg. \bigg]
    \end{aligned}
\end{equation}
\begin{equation}\label{eq30}
    \begin{aligned}
    E_{diss,a,\perp} &= 2\pi\sigma_{0}\times \operatorname{Im}\left[\frac{\alpha_{f_1}\theta_{1,f_1}}{\gamma_{f_1}}\sinh\left(\frac{\gamma_{f_1}t_{1}}{2} \right) \right. \\
    &\quad+\frac{2\nu_{f_1}}{1-\nu_{f_1}}\alpha_{s}\theta_{1,s}\cosh(\gamma_{s}t_{s})\times\frac{t_{1}}{2}\\
    &\quad +\frac{\alpha_{f_2}\theta_{1,f_2}}{\gamma_{f_2}}\sinh\left(\frac{\gamma_{f_2}t_{2}}{2} \right) \\
    &\quad +\frac{2\nu_{f_2}}{1-\nu_{f_2}}\alpha_{s}\theta_{1,s}\cosh(\gamma_{s}t_{s})\times\frac{t_{2}}{2}\bigg. \bigg]
    \end{aligned}
\end{equation}

From elasticity theory, the elastic energy stored in the case of parallel and perpendicular fields, respectively, in these half layers can be expressed as
\begin{equation}\label{eq31}
    \begin{aligned}
    E_{stored,a,\parallel}&=\frac{1}{2}\sigma_0^{2}\frac{t_{1}}{2}\times\frac{2(1-\nu_{f_1})}{E_{f_1}} \\
    &\quad + \frac{1}{2}\sigma_0^{2}\frac{t_{2}}{2}\times\frac{2(1-\nu_{f_2})}{E_{f_2}} \text{ ,} \\
    E_{stored,a,\perp}&=\frac{1}{2}\sigma_0^{2}\frac{t_{1}}{2}\times\frac{(1-2\nu_{f_{1}})(1+\nu_{f_{1}})}{E_{f_{1}}(1-\nu_{f_{1}})} \\
    &\quad + \frac{1}{2}\sigma_0^{2}\frac{t_{2}}{2}\times\frac{(1-2\nu_{f_{2}})(1+\nu_{f_{2}})}{E_{f_{2}}(1-\nu_{f_{2}})}
    \end{aligned}
\end{equation}

\textcolor{black}{For the thermoelastic loss due to heat conduction between the last layer of Material $2$ and the substrate [illustrated in Fig.~\ref{multilayer_model}(b)], indicated by subscript $b$ and by adopting the boundary conditions for single-layer coated substrates, we find the energy dissipated in the form of}
\begin{equation}\label{eq32}
    \begin{aligned}
    E_{diss,b,\parallel} &= 2\pi\sigma_{0}\times \operatorname{Im}\left[(2\nu_{f_2}-2) \right. \\
    &\quad \times\frac{\alpha_{f_2}\theta^{'}_{1,f_2}}{\gamma_{f_2}}\sinh\left(\frac{\gamma_{f_2}t_{2}}{2}\right) \\
    &\quad +(4-2\nu_{f2})\alpha_{s}\theta_{1,s}\cosh(\gamma_{s}t_{s})\times\frac{t_{2}}{2}\\
    &\quad -\frac{2\alpha_{s}\theta_{1,s}}{\gamma_{s}}\sinh(\gamma_{s}t_{s})\bigg. \bigg]
    \end{aligned}
\end{equation}
\begin{equation}\label{eq33}
    \begin{aligned}
    E_{diss,b,\perp} &= 2\pi\sigma_{0}\times \operatorname{Im}\left[\frac{\alpha_{f_2}\theta^{'}_{1,f_2}}{\gamma_{f_2}}\sinh\left(\frac{\gamma_{f_2}t_{2}}{2}\right) \right. \\
    &\quad +\frac{2\nu_{f_2}}{1-\nu_{f_2}}\alpha_{s}\theta_{1,s}\cosh(\gamma_{s}t_{s})\times \frac{t_{2}}{2}\\
    &\quad -\frac{1+\nu_{s}}{1-\nu_{s}}\frac{\alpha_{s}\theta_{1,s}}{\gamma_{s}}\sinh(\gamma_{s}t_{s})\bigg. \bigg]
    \end{aligned}
\end{equation}
where $\theta^{'}_{1,f_{2}}$ and $\theta_{1,s}$ are defined as
\begin{equation}\label{eq34}
    \begin{aligned}
    \theta^{'}_{1,f_{2}} &= \frac{\Delta\beta^{''}}{\cosh\left(\frac{\gamma_{f_{2}}t_{2}}{2}\right)+R^{''}\sinh\left(\frac{\gamma_{f_{2}}t_{2}}{2}\right)\coth(\gamma_{s}t_{s})} \text{ ,} \\
    \theta_{1,s} &=
    -\frac{\Delta\beta^{''} R^{''}}{\coth\left(\frac{\gamma_{f_{2}}t_{2}}{2}\right)\sinh(\gamma_{s}t_{s})+R^{''}\cosh({\gamma_{s}t_{s})}}
    \end{aligned}
\end{equation}
and where $\Delta\beta^{''} = \beta_{f_{2}}-\beta_{s}$ and $R^{''}=(k_{f_{2}}\gamma_{f_{2}})/(k_{s}\gamma_{s})=\sqrt{(k_{f_{2}}C_{f_{2}})/(k_{s}C_{s})}$.\\

The energy stored for parallel and perpendicular fields is expressed in the form of
\begin{equation}\label{eq35}
    \begin{aligned}
    E_{stored,b,\parallel}&=\frac{1}{2}\sigma_0^{2}
    \frac{t_{2}}{2}\times\frac{2(1-\nu_{f_{2}})}{E_{f_{2}}} \\
    &\quad + \frac{1}{2}\sigma_0^{2}t_{s}\times\frac{2(1-\nu_{s})}{E_{s}} \\ 
    E_{stored,b,\perp}&=\frac{1}{2}\sigma_0^{2} \frac{t_{2}}{2}\times\frac{(1-2\nu_{f_{2}})(1+\nu_{f_{2}})}{E_{f_{2}}(1-\nu_{f_{2}})} \\
    &\quad + \frac{1}{2}\sigma_0^{2}t_{s}\times\frac{(1-2\nu_{s})(1+\nu_{s})}{E_{s}(1-\nu_{s})}
    \end{aligned}
\end{equation}

The $E_{diss}$ and $E_{stored}$ terms in the first deposited layer of Material $2$ are double counted in the two regions. Therefore, the expression of the total thermoelastic loss of a homogeneous system, a substrate coated with a multilayer, is given by
\begin{equation}\label{eq36}
    \phi = \frac{E_{diss,a}\times N-E_{diss,a_{2}}+E_{diss,b}}{2\pi\times(E_{stored,a}\times N-E_{stored,a_{2}}+E_{stored,b})}
\end{equation}
where $N$ is the total number of layers, $E_{diss,a_{2}}$ and $E_{stored,a_{2}}$ stand for the energy dissipated and stored in half a layer of Material 2, correspondingly.

\section{Results and Discussion}\label{results}
In this section, we present the thermoelastic loss $\phi$ due to thermal expansion mismatch between different materials for various cases, which is calculated using the equations presented in the previous sections. The input elastic field, unless specified, uses the normal stress field described in Section~\ref{appliedelasticfield}, as it is commonly seen in many applications and used in the measurement of coating properties. The physical properties of the materials used for these calculations are listed in Table~\ref{table1}. We note that while small variations in the elastic properties, up to 30\%, do not have a significant effect in the accuracy of the thermoelastic loss calculations, variations in the thermal properties have a much larger effect. Therefore, accurate values of the thermal properties are necessary to obtain reliable estimations of the thermoelastic loss.

\begin{table*}[!h]
\renewcommand{\arraystretch}{1.2}
\footnotesize
\centering
\caption{Elastic and thermal properties at room temperature of materials used for the calculations of $\phi$: Young's modulus $E$, Poisson's ratio $\nu$, thermal expansion coefficient $\alpha$, specific heat per unit volume $C_V$, and thermal conductivity $k$. The temperature dependencies of the thermal properties ($\alpha$, $C_V$ and $k$) are implemented for the calculations of $\phi$ as reported in their corresponding references.\\}
\begin{threeparttable}
\setlength{\tabcolsep}{5pt}
\begin{tabular}{lccccc} \hline \hline
    Materials & $E$ & $\nu$ & $\alpha$ & $C_V$ & $k$ \\
    & (GPa) &  & (10$^{-6}$ K$^{-1}$) & (10$^{6}$ J\,m$^{-3}$\,K$^{-1}$) & (W\,m$^{-1}$\,K$^{-1}$) \\ \hline
    Fused $a$-SiO$_2$ & 71~\cite{COMTE200242} & 0.17~\cite{COMTE200242} & 0.6~\cite{KUHN2009323} & 1.6~\cite{RICHET19822639} & 1.4~\cite{10.1063/1.4764904} \\
    c-Si & 169~\cite{Hopcroft2010WhatSilicon} & 0.28~\cite{Hopcroft2010WhatSilicon} & 2.6~\cite{White1997ThermophysicalUpdate} & 1.6~\cite{Flubacher1959TheSpectra} & 92.0~\cite{Glassbrenner1964ThermalPoint} \\
    $a$-Si & 115\,\tnote{a} & 0.18\,\tnote{a} & 2.4\,\tnote{b}~~\cite{DeLima1999CoefficientFilms} & 2.4~\cite{Queen2013ExcessSilicon} & 91.0~\cite{Zink2006ThermalSilicon} \\
    $a$-SiO$_2$ & 70~\cite{Granata2020} & 0.19~\cite{Granata2020} & 0.6\,\tnote{c}~~\cite{El-Kareh1995FundamentalsTechnology} & 2.2\,\tnote{c}~~\cite{El-Kareh1995FundamentalsTechnology} & 1.2\,\tnote{c}~~\cite{El-Kareh1995FundamentalsTechnology} \\
    $a$-Ti:Ta$_2$O$_5$ & 120~\cite{Granata2020} & 0.29~\cite{Granata2020} & 3.9~\cite{Abernathy2014InvestigationFilms} & 2.1~\cite{Fejer2004ThermoelasticDetectors} & 33~\cite{Fejer2004ThermoelasticDetectors}\\
    $c$-GaAs & 85.5~\cite{Sze2006AppendixGaAs} & 0.31~\cite{Sze2006AppendixGaAs} & 5.75~\cite{Sze2006AppendixGaAs} & 1.74~\cite{Sze2006AppendixGaAs} & 46~\cite{Sze2006AppendixGaAs}\\
    $c$-Al$_{0.92}$Ga$_{0.08}$As & 100~\cite{Chalermsongsak2016CoherentMirrors} & 0.32~\cite{Chalermsongsak2016CoherentMirrors}  & 5.2~\cite{Chalermsongsak2016CoherentMirrors}  & 1.7~\cite{Chalermsongsak2016CoherentMirrors}  & 70~\cite{Chalermsongsak2016CoherentMirrors}\\
    \hline \hline
\end{tabular}
\begin{tablenotes}
\footnotesize
    \item[a] $a$-Si elastic properties are obtained using $E=2G(1+\nu)$, where $E=(1-\nu)140$ GPa~\cite{Witvrouw1993ViscosityGe}, and the shear modulus $G=49$ GPa~\cite{Molina-Ruiz2021OriginSilicon}, both experimentally measured.
    \item[b] $a$-Si thermal expansion coefficient temperature dependence is assumed to be the same than for c-Si reported in Ref.~\cite{White1997ThermophysicalUpdate}.
    \item[c] $a$-SiO$_2$ thermal properties temperature dependence is assumed to be the same as for fused $a$-SiO$_2$ reported in Refs.~\cite{Jacobs1984ThermalMirrors, Zeller1971ThermalSolids}.
\end{tablenotes}
\end{threeparttable}
\label{table1}
\end{table*}

\subsection{Thickness and frequency dependence}
\begin{figure}[!h]
    \centering
    \includegraphics[width=8.0cm]{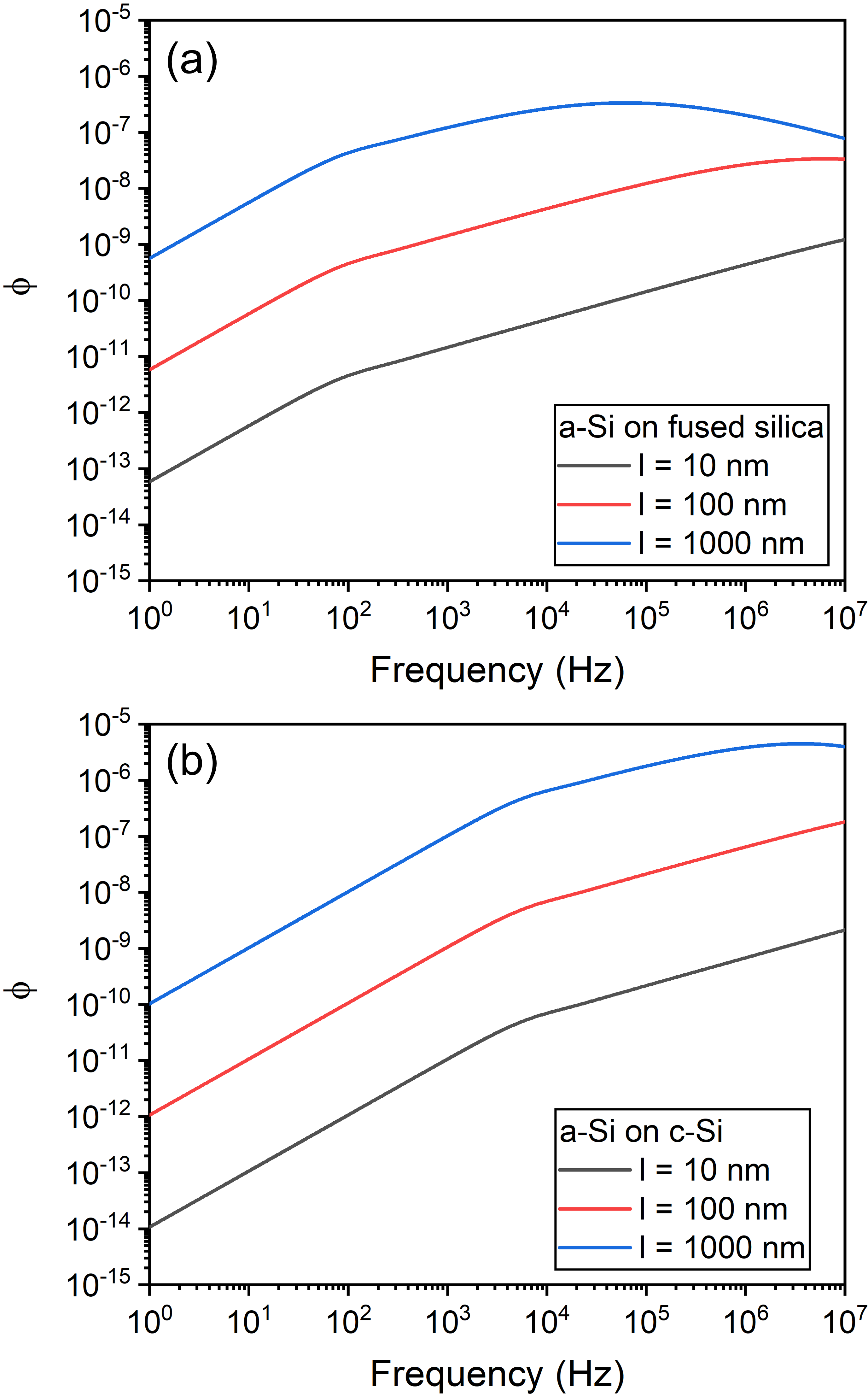}
    \caption{Thermoelastic loss $\phi$ as a function of frequency calculated at room temperature for an $a$-Si layer 10-nm-thick (black line), 100-nm-thick (red line) and 1,000-nm-thick (blue line) onto 100-$\upmu$m-thick substrates: (a) fused silica (a-SiO$_2$) and (b) crystalline silicon (c-Si).}
    \label{filmthicknessthintopsilicabottomcsi}
\end{figure}

The thermoelastic loss of single-layer coated substrates was first investigated. $\phi$ of amorphous silicon ($a$-Si) films of various thicknesses, deposited on fused silica and crystalline Si (c-Si) substrates at room temperature (RT) is plotted as a function of frequency in Fig.~\ref{filmthicknessthintopsilicabottomcsi}. The numerical values of $\phi$ are obtained from Eq.~\ref{eq25}, for a frequency range from $1$ to $10^7$ Hz. In Fig.~\ref{filmthicknessthintopsilicabottomcsi}, $\phi$ shows a non-monotonic behavior with one or two inflection points as a function of frequency. The thermoelastic loss exhibits a Lorentzian behavior as a function of the vibration frequency $\omega$ and the relaxation rate $\tau$ of the system, with a maximum value when $\omega\tau = 1$~\cite{Lifshitz2000ThermoelasticSystems}. The inflection points shown in Fig.~\ref{filmthicknessthintopsilicabottomcsi} are, in fact, $\phi$ local maxima, dominated by either the coating or the substrate, and occur at a frequency $f_{lm} = 1 / (2\pi \tau)$, where $\tau = \ell^2 C_V / k$ is the thermal diffusion time through a layer of thickness $\ell$. In Fig.~\ref{filmthicknessthintopsilicabottomcsi}, the inflection points at low and high frequencies are dominated by the substrate and by the coating, respectively. The larger the thermal diffusion time through the coating or the substrate, the lower the frequency at which the corresponding inflection point will appear.

\begin{figure}[!h]
    \centering
    \includegraphics[width=8.0cm]{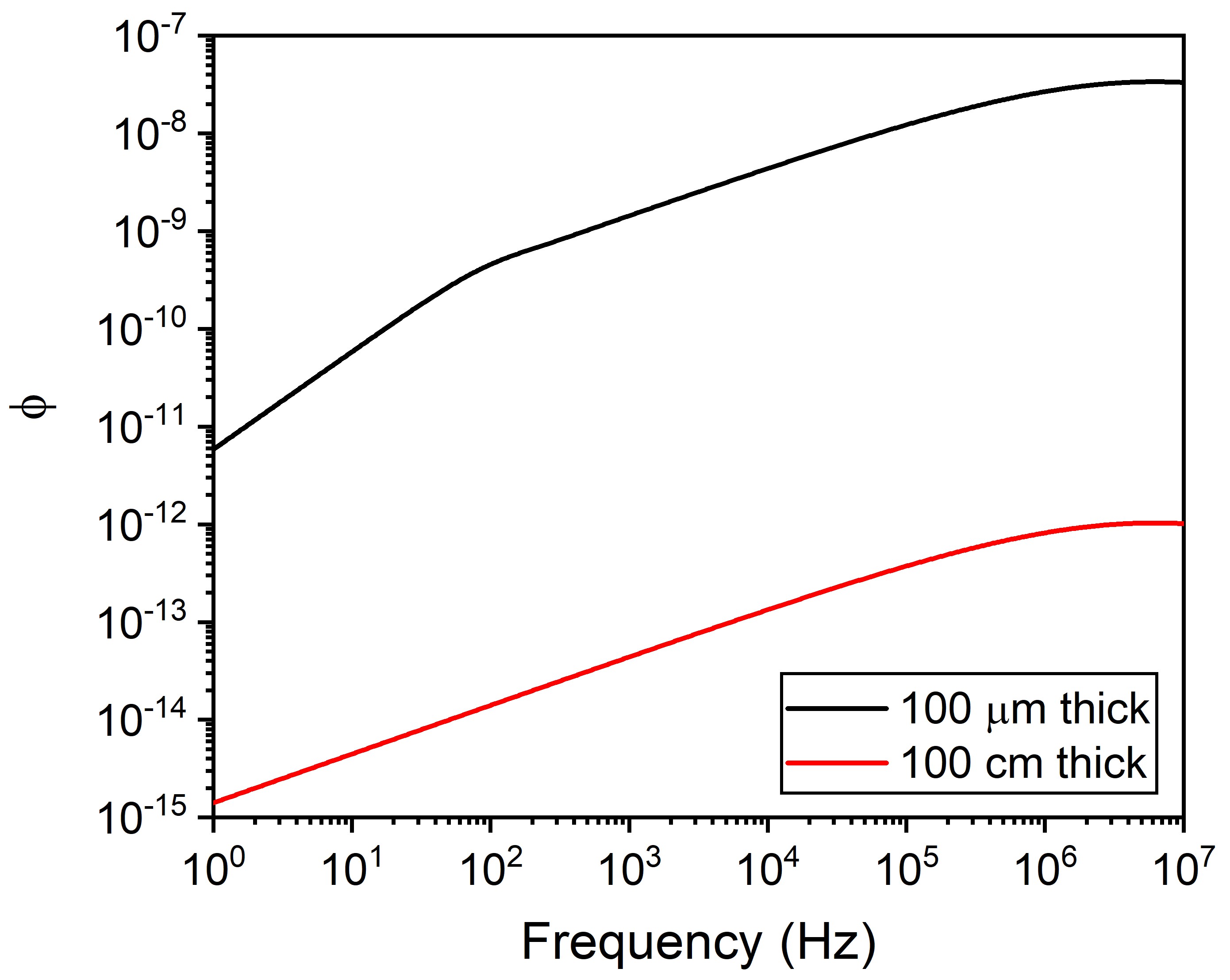}
    \caption{Thermoelastic loss $\phi$ as a function of frequency calculated at room temperature for a 100-nm-thick $a$-Si layer onto 100-$\upmu$m-thick (black line) and 100-cm-thick (red line) silica substrates.}
    \label{silicasubstratethickness}
\end{figure}

These results show that a thicker film would yield higher thermoelastic loss, which is a direct outcome of a larger temperature difference, and thus a higher heat flux between the film and the substrate. The substrate thickness also has an impact on $\phi$; thicker substrates lead to an approximately proportional reduction in the thermoelastic loss, which is demonstrated in Fig.~\ref{silicasubstratethickness}. As discussed at the end of Section~\ref{energydissipation}, while the elastic energy stored is linearly proportional to the substrate thickness, there is only a marginally small increase in the overall energy dissipation, which results in near proportional damping. Eventually, $\phi$ would approach zero at infinite substrate thickness, as the energy loss caused by thermal expansion mismatch would be vanishingly small in comparison to the elastic energy stored in the substrate.

Considering that both film and substrate thicknesses have an impact on the thermoelastic loss, we note that certain film to substrate thickness ratios may cause the thermoelastic loss to be one of the main energy dissipation mechanisms in the system. These results highlight a remarkable difference with other dissipation mechanisms, such as mechanical loss and thermoelastic loss due to statistical fluctuations, where the loss factor is intrinsic to the material and independent of the system's volume. Themoelastic loss due to thermal expansion mismatch depends on the system's volume.

\subsection{Effect of input stress field}
\begin{figure}[!h]
    \centering
    \includegraphics[width=8.0cm]{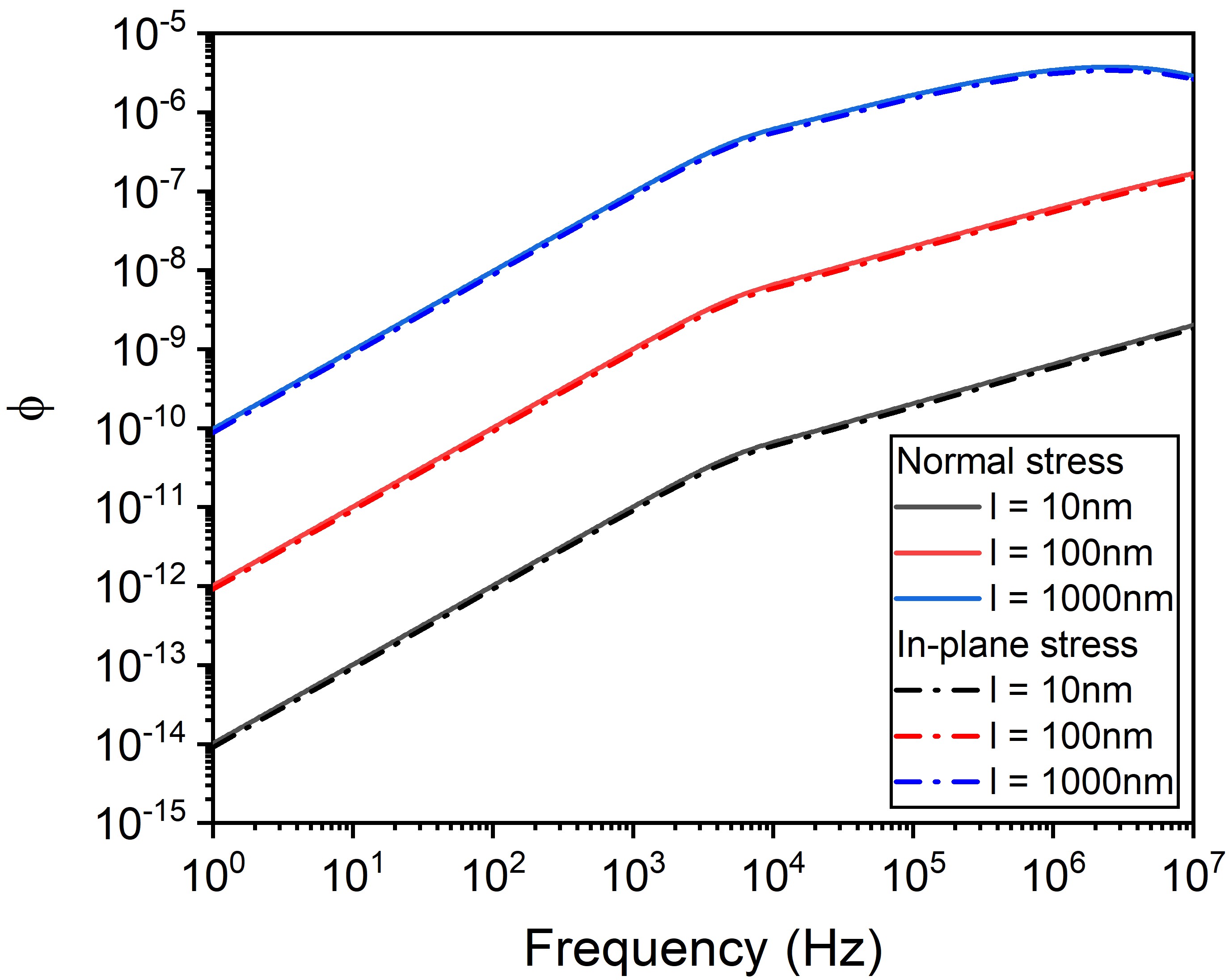}
    \caption{Thermoelastic loss $\phi$ as a function of frequency calculated at room temperature for a 100-$\upmu$m-thick c-Si substrate coated with an $a$-Si layer 10-nm-thick (black lines), 100-nm-thick (red lines) and 1,000-nm-thick (blue lines) and with either normal stress (solid lines) or in-plane stress (dashed lines).}
    \label{stressfield}
\end{figure}

As previously mentioned in Section~\ref{appliedelasticfield}, another common type of stress is in-plane stress. Figure~\ref{stressfield} compares the thermoelastic loss for the in-plane and perpendicular stress components. The inflection points position and intensity are only slightly affected by the field polarization; the difference is within $3\%$ and $11\%$, respectively. Other types of stress (bulk, shear...) can also be incorporated into the calculation, which makes this model a powerful tool to perform calculations in complex systems.

\subsection{Multilayer coating}
\begin{figure}[!h]
    \centering
    \includegraphics[width=8.0cm]{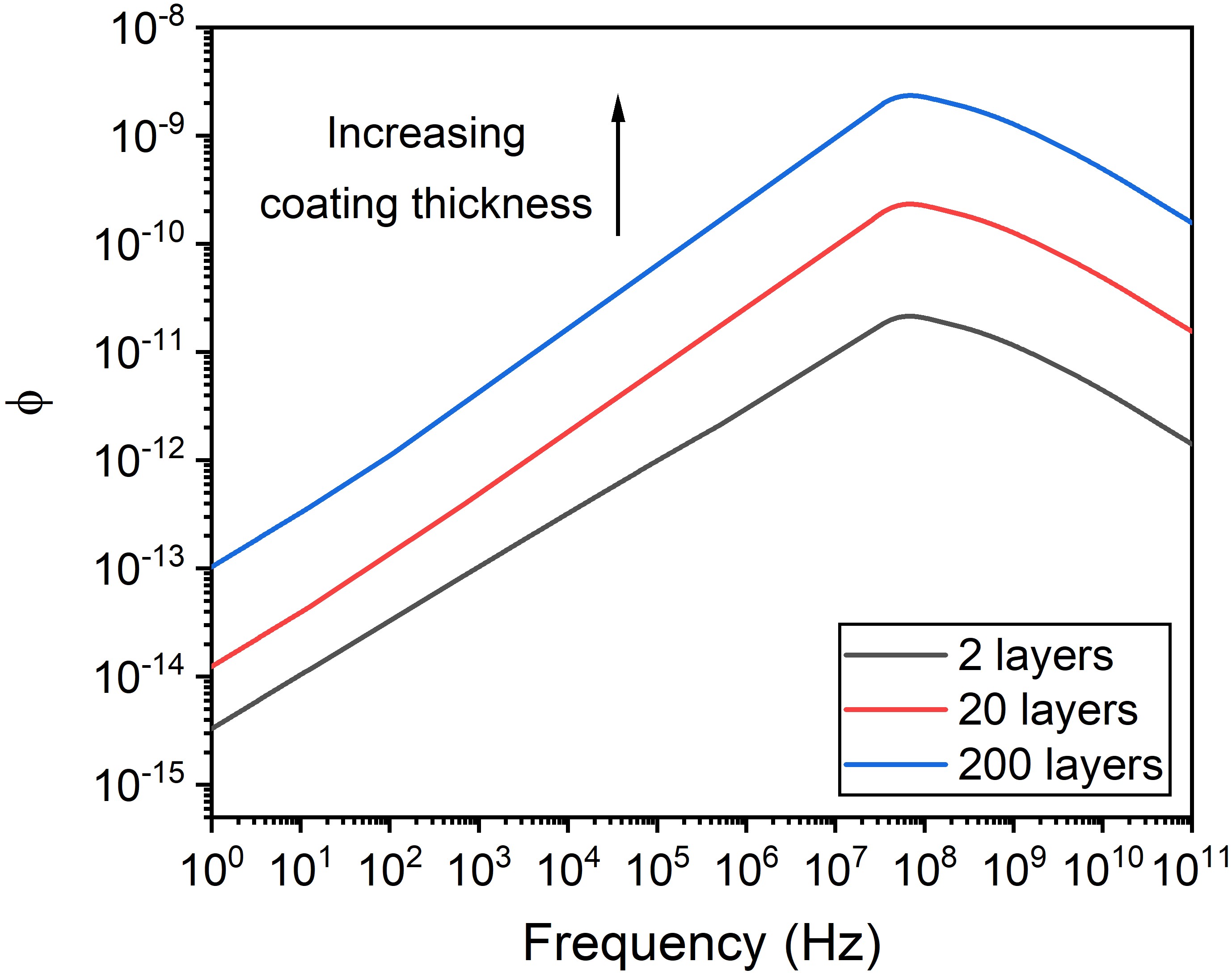}
    \caption{Thermoelastic loss $\phi$ as a function of frequency calculated at room temperature for a 10-cm-thick fused silica substrate coated with a variable number of alternating layers of $a$-Si (76.0 nm) and $a$-SiO$_2$ (184.7 nm): 2 layers (black line), 20 layers (red line) and 200 layers (blue line). We used $n=3.5$ for $a$-Si~\cite{Pierce1972ElectronicStudies} and $n=1.44$ for $a$-SiO$_2$~\cite{Leviton2006Temperature-dependentSilica}.}
    \label{multi_nooflayers}
\end{figure}

We now consider a multilayer coating consisting of amorphous silicon and amorphous silica on top of a fused silica substrate. The substrate thickness is set to be $100$ $\mu$m and the layer thicknesses are selected to follow the quarter-wavelength rule, $\lambda/4n$, where $\lambda$ is the wavelength of the operating laser and $n$ is the refractive index~\cite{Sheppard1995ApproximateMedium}. We first investigated the effect of the number of layers. As can be seen in Fig.~\ref{multi_nooflayers}, while the peak position does not change noticeably, the thermoelastic loss increases in proportion to the number of layers with a constant thickness, i.e., $\phi$ increases proportional to the amount of material.

\begin{figure}[!h]
    \centering
    \includegraphics[width=8.0cm]{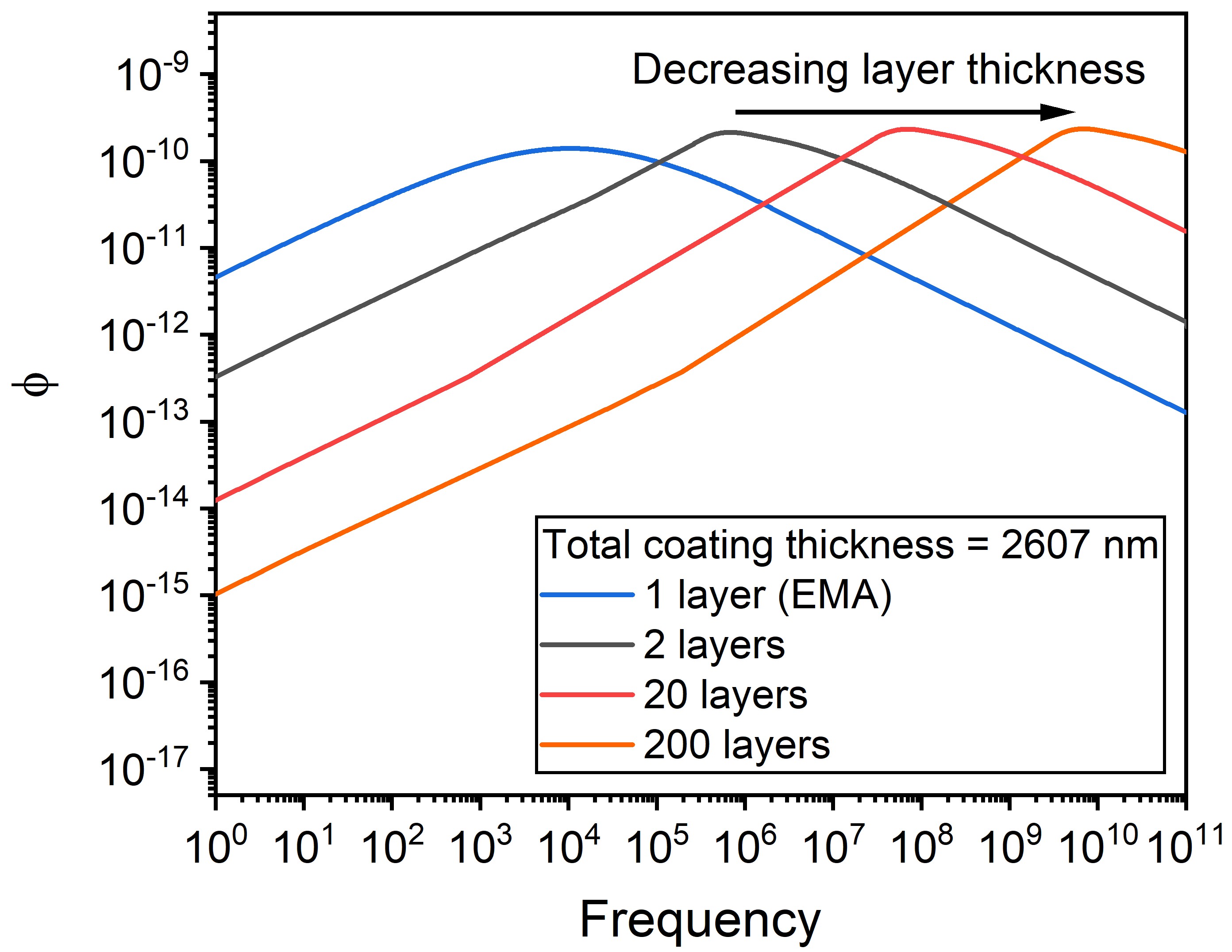}
    \caption{Thermoelastic loss $\phi$ as a function of frequency calculated at room temperature for a 10-cm-thick fused silica substrate coated with a variable number of alternating layers of $a$-Si and silica keeping a coating total thickness of $2607$ nm: 1 layer using the EMA (blue line), 2 layers consisting of a 760-nm-thick $a$-Si layer and a 1847-nm-thick $a$-SiO$_2$ layer (black line), 20 layers consisting of 76.0-nm-thick $a$-Si layers and 184.7-nm-thick $a$-SiO$_2$ layers (red line), and 200 layers consisting of 7.60-nm-thick $a$-Si layers and 18.47-nm-thick $a$-SiO$_2$ layers (orange line).}
    \label{multi_comparison_layerthickness}
\end{figure}

The effective medium approach EMA considers the two materials used in the multilayer coating as one homogeneous film with averaged physical properties. In contrast, our method treats each layer as an individual component and computes the loss generated from heat transport across all interfaces. In Fig.~\ref{multi_comparison_layerthickness} we compare the results obtained by our model and the EMA. The total film thickness is kept constant while the thickness of the layers varies. It can be seen that, the $\phi$ peaks are approximately at the same loss intensity, which is expected since $\phi$ is proportional to the total volume of the system. However, our model predicts that the number of interfaces plays a role and affects the loss peak position. This result is a consequence of the layers' thickness effect on the system's thermal field that modifies the coating temperature profile as depicted in Fig.~\ref{multilayer_model}. This effect is implicit in the frequency dependence of the thermal field (see Section~\ref{formulationthermalfield}) and can be observed comparing Figs.~\ref{multi_nooflayers} and \ref{multi_comparison_layerthickness}; the shift in frequency of the thermoelastic loss peak is a consequence only of the layers' thickness, not their number.

Figure~\ref{multi_comparison_layerthickness} shows that by reducing the thickness of the layers at constant coating thickness, i.e., increasing the number of interfaces, the loss peak position is shifted towards higher frequencies, which effectively lowers $\phi$ in the frequency range of interest for GW detectors (10 Hz to 10 kHz), such as LIGO, Virgo and the Kamioka Gravitational-Wave detector (KAGRA)~\cite{Aso2013InterferometerDetector}. Our model can analytically estimate $\phi$ for any layer thickness, even nanolayered coatings~\cite{Pan2014, Magnozzi2018, Kuo2019}. However, the model predictions will remain valid only if the system components, its substrate and layers, are homogeneous and their thermal and elastic properties are known.

\subsection{Comparison of mirror coatings for gravitational-wave detectors}
Based on the model presented in Section~\ref{multilayer}, the thermoelastic loss of multilayer coating is calculated for various combinations of materials and layer thicknesses at three excitation frequencies, 100 Hz, 1 kHz and 10 kHz. The predicted values are listed in Table~\ref{table2}. Our calculation shows that the choice of mirror coating and substrate in Advanced LIGO (the current LIGO generation); using a coating made of amorphous titania-doped tantala and amorphous silica and a fused silica substrate, has the lowest thermoelastic loss at all three frequencies compared to other candidate materials.

\begin{table*}[t]
\renewcommand{\arraystretch}{1.2}
\footnotesize
\centering
\caption{Estimated thermoelastic loss $\phi$ at different frequencies for current Advanced LIGO mirrors and potential designs at different laser wavelengths $\lambda$. All $\phi$ values are calculated at room temperature, except those of the silicon-silica coating, which are estimated at 123 K. The composition and thickness details of each mirror design are shown at the bottom of the Table as [thickness layer 1 / thickness layer 2]$_\text{(layers number)}$ // thickness substrate.\\}
\begin{threeparttable}
\setlength{\tabcolsep}{12pt}
\begin{tabular}{lcccc} 
    \hline \hline
    Mirror design & $\lambda$ & \multicolumn{3}{c}{$\phi$}\\
     & nm & 100 Hz & 1 kHz & 10 kHz\\ \hline
    Advanced LIGO\tnote{1} & 1064 & $2.8\times10^{-13}$  & $9.4\times10^{-13}$ & $3.2\times10^{-12}$\\
    & & & & \\
    Multimaterial coating\tnote{2} & 1064 & $8.4\times10^{-12}$ & $2.7\times10^{-11}$ & $9.4\times10^{-11}$\\
    & & & & \\
    \multirow{2}{*}{Silicon-silica coating\tnote{3}} & 1550 & $5.4\times10^{-22}$ & $3.4\times10^{-21}$ & $2.7\times10^{-20}$\\
    & 2000 & $1.0\times10^{-21}$ & $6.8\times10^{-21}$ & $5.3\times10^{-20}$\\
    & & & & \\
    Crystalline coating\tnote{4} & 1064 & $6.2\times10^{-12}$ & $2.0\times10^{-11}$ & $4.6\times10^{-11}$\\
    \hline \hline
\end{tabular}
\begin{tablenotes}
\footnotesize
    \item[1] \cite{Aasi2015AdvancedLIGO}~\,: [128.5 nm $a$-Ti:Ta$_2$O$_5$ / 183.4 nm $a$-SiO$_2$]$_{(30)}$ // 10 cm $a$-SiO$_2$
    \item[2] \cite{Steinlechner2015ThermalCoatings}: [176.0 nm $a$-Ta$_2$O$_5$ or 111.0 nm $a$-Si / 267.0 nm $a$-SiO$_2$]$_{(30)}$ // 10 cm $a$-SiO$_2$
    \item[3] \cite{Adhikari2020ADetection}: [76.0 nm $a$-Si / 184.7 nm $a$-SiO$_2$]$_{(16)}$ // 55 cm c-Si
    \item[4] \cite{Koch:19}: [76.4 nm $c$-GaAs / 89.4 nm $c$-AlGaAs]$_{(70)}$ // 10 cm $c$-Si
\end{tablenotes}
\end{threeparttable}
\label{table2}
\end{table*}

\subsection{Dependence on temperature}
\begin{figure}[!h]
    \centering
    \includegraphics[width=8.0cm]{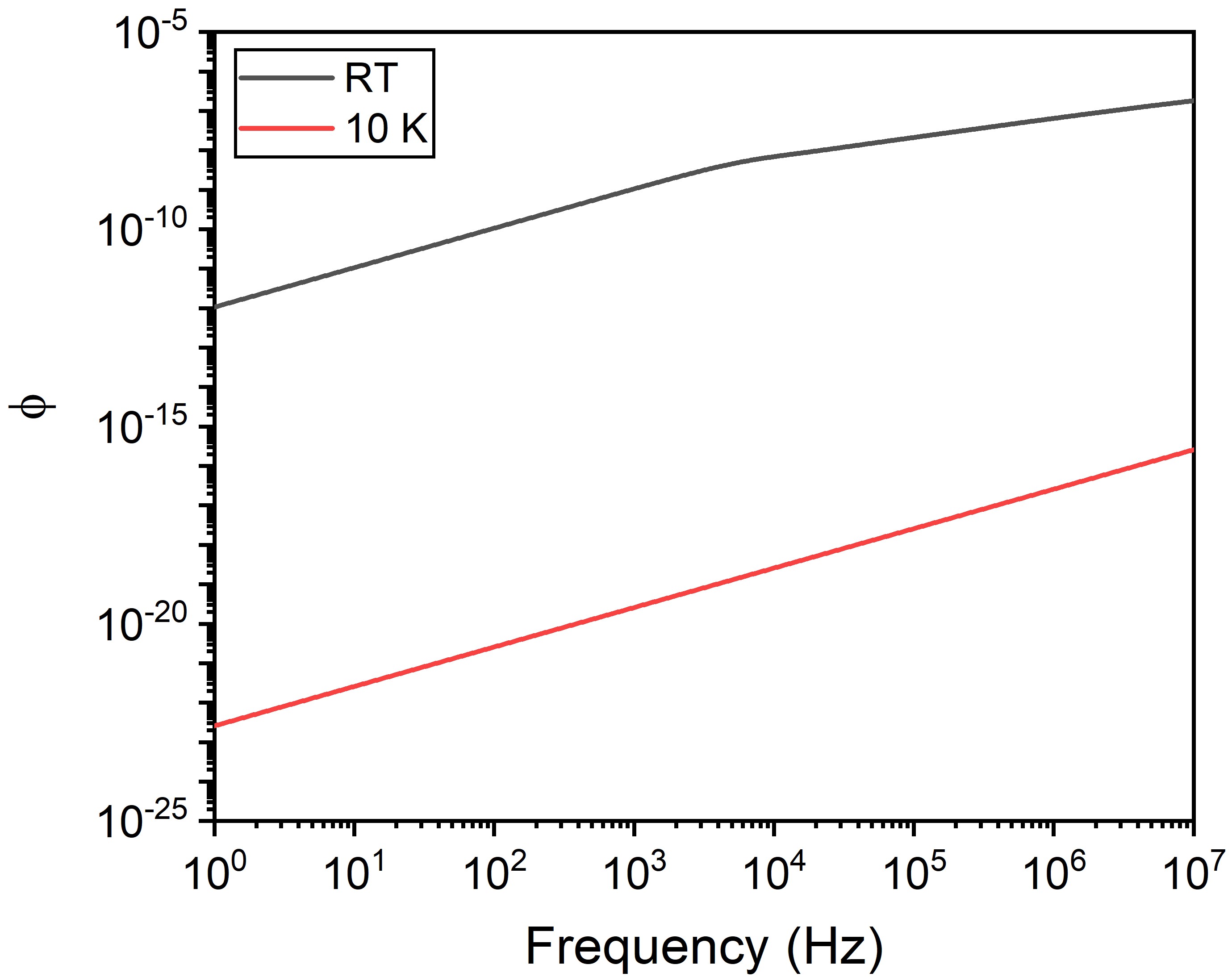}
    \caption{Thermoelastic loss $\phi$ as a function of frequency calculated at room temperature RT (black line) and at 10 K (red line) for a c-Si substrate 100-$\upmu$m-thick coated with a 100-nm-thick $a$-Si layer.}
    \label{lowTcomparison}
\end{figure}

Thermoelastic loss is largely dependent on temperature T since the induced thermal field is a function of T, as seen in Eq.~\ref{eq1}. In addition, the thermal properties of materials (coefficient of thermal expansion $\alpha$, specific heat $C_V$, and thermal conductivity $k$) are temperature dependent. As a result, $\phi$ varies with temperature. In this Section, we consider that the elastic properties do not change with temperature since their dependence with T is much weaker than that of the thermal properties. We calculate $\phi(T)$ using experimental values of the thermal and elastic properties of the materials used in the system, and when those are not available, we make reasonable estimations. We assume that amorphous silicon has the same thermal expansion coefficient of its crystalline form~\cite{White1997ThermophysicalUpdate}, while its thermal conductivity and heat capacity are taken from Refs.~\cite{Zink2006ThermalSilicon,Queen2013ExcessSilicon}. The values for the thermal properties of fused silica are taken from Refs.~\cite{Fukuhara1997LowQuartz, Jacobs1984ThermalMirrors, Zeller1971ThermalSolids}. Figure~\ref{lowTcomparison} illustrates that the thermoelastic loss of 100 nm-thick $a$-Si onto a silica substrate 100 $\mu$m-thick is significantly lower at 10 K than at RT for all frequencies, and the inflection points are shifted towards higher frequencies.

The thermal expansion coefficient of crystalline silicon $\alpha_{c-Si}$ is zero at $17.6$ K and $123.7$ K~\cite{White1997ThermophysicalUpdate,Middelmann2015ThermalK}. For this reason, future cryogenic GW detectors, such as LIGO Voyager, plan to operate at $123.7$ K using silicon-based mirrors to eliminate the thermoelastic loss due to thermal mismatch between different materials~\cite{Adhikari2020ADetection}. $a$-Si has good mechanical loss and optical reflectivity compared to the currently used $a$-Ti:Ta$_2$O$_5$ layer in LIGO and Virgo~\cite{Murray2015Ion-beamSystems, Birney2018AmorphousAstronomy, Steinlechner2021HowDetectors}. Future LIGO Voyager plans to use crystalline silicon for the mirror substrate, and amorphous silicon as the high-index material and silica as the low-index layer for the mirror coating. This detector will operate at cryogenic temperatures to further reduce the loss and improve the sensitivity beyond the detection limits of the current GW detectors. 

\begin{figure}[!h]
    \centering
    \includegraphics[width=8.0cm]{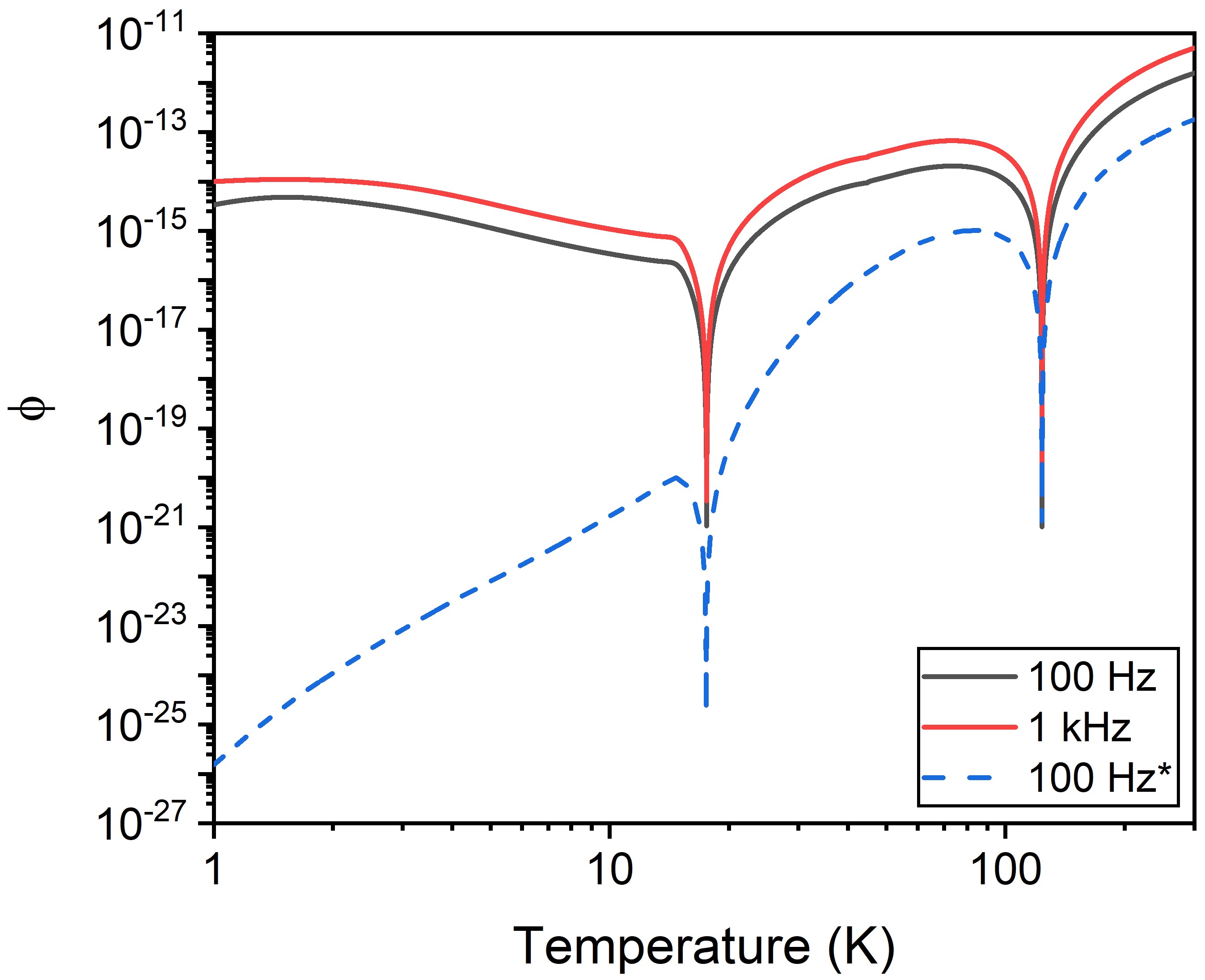}
    \caption{Thermoelastic loss $\phi$ as a function of temperature for a 55-cm-thick c-Si substrate coated with 16 alternating layers of $a$-Si (142.9 nm) and $a$-SiO$_2$ (347.2 nm). $\phi$ is calculated at an excitation frequency of 100 Hz (black line) and 1 kHz (red line). The 100 Hz* data (blue dashed line) is obtained reversing the order of the layers in the coating so that $a$-Si is in contact with the c-Si substrate, which highlights the effect of minimizing the thermal expansion mismatch between the substrate and the layer in contact with the coating.}
    \label{multicryof}
\end{figure}

We calculated the thermoelastic loss of the proposed multilayer stack of $a$-Si/$a$-SiO$_2$ films for the LIGO Voyager mirror coating at two different excitation frequencies. As plotted in Figure \ref{multicryof}, $\phi$ increases with increasing the vibration frequency, in agreement with the results previously discussed, and shows two deep minima when the thermal expansion coefficient $\alpha$ of c-Si is zero; at 17.6 and 123.7 K~\cite{White1997ThermophysicalUpdate}. For this calculation we assumed that $\alpha$ of $a$-Si is the same as that of c-Si.

The predicted loss shown in Figure \ref{multicryof} is based on the structure where $a$-SiO$_2$ is the first layer and in contact with the substrate. We note that when $a$-Si is the first layer, the thermoelastic loss is significantly lower at all temperatures (blue dashed line in Fig.~\ref{multicryof}). This happens because the thermal expansion mismatch between $a$-Si and c-Si is significantly smaller than that between $a$-SiO$_2$ and c-Si. If the thermal expansion coefficient of $a$-Si is not exactly zero at the same temperatures as for $c$-Si (17.6 K and 123.7 K), then the two minima seen in Fig.~\ref{multicryof} would split into four, one to the left and one to the right of the current peaks. This could affect the current plans of making 123.7 K the operating temperature for LIGO Voyager.

\section{Conclusions}
We present a mathematical model to calculate the thermoelastic loss due to thermal expansion mismatch between different materials. The results obtained highlight how material properties, measurement temperature and frequency, and mirror design (materials, thickness of layers and number of interfaces) affect the thermoelastic loss and, therefore, the thermal noise that limits the resolution in sensing applications. In the frequency range studied, thermoelastic loss increases with frequency up to a maximum value, related to the thermal diffusion time through the coating, and decreases above that.

For a single-layer coating, we find that thermoelastic loss increases with the coating thickness, and reduces with a thicker substrate for a fixed coating thickness. We extend our analytical solutions to multilayered structures, specifically without using the effective medium approach, which therefore allows us to calculate the effect of interfaces due to the mismatch of thermal expansion between neighboring layers. We demonstrate that thermoelastic loss correlates with the number of layers, or interfaces, for a given total thickness. At constant coating thickness, the thermoelastic loss curve shifts toward higher frequencies when the number of layers, or interfaces, increase, which therefore implies that thermoelastic loss decreases in the frequency range relevant for gravitational-wave detectors. We show that thermoelastic loss is proportional to temperature due to the dependence of thermal field and materials properties with temperature.

We show that future mirrors should consider the thermal expansion of the materials used and their mismatch, including in particular the material chosen for the first layer in contact with the substrate, the thickness of layers and the number of interfaces within the coating.

\vspace{0.5 cm}

%\begin{acknowledgments}
We gratefully thank M. M. Fejer for fruitful discussions, and the support of the LIGO Scientific Collaboration Center for Coatings Research, jointly funded by the United States National Science Foundation (NSF) and the Gordon and Betty Moore Foundation through Grant No. 6793. We also thank the support of the NSF through Grant No. DMR-1809498.
%\end{acknowledgments}

\appendix

\section{Thermal field solutions}
\label{sec:appendix-thermalfields}
As discussed in Section~\ref{formulationthermalfield}, two components are required for the expression of the thermal fields in the $z$-direction, the particular solution and the specific solution. The particular solution is assumed not to depend on the in-plane position if both film and substrate are homogeneous. Looking at Eq.~\ref{eq2}, we can conclude that
\begin{equation}\label{A1}
    \theta_{p,j} = -\beta_j
\end{equation}
Using the boundary conditions, we can solve for the specific solution
\begin{equation}\label{A2}
    \begin{aligned}
    \theta_{s,f} &= \theta_{1,f}\cosh(\gamma_f z) \text{ ,} \\
    \theta_{s,s} &= \theta_{1,s}\cosh[\gamma_s (h-z)]
    \end{aligned}
\end{equation}
and $\gamma_j = (1+i)\sqrt{\omega/(2\kappa_j)}=(1+i)\sqrt{\pi f C_{j}/k_{j}}$, so that
\begin{equation}\label{A3}
    \begin{aligned}
    \frac{\partial\theta_{p,j}}{\partial t}-\kappa_j\frac{\partial^2\theta_{p,j}}{\partial z^2} &= -i\omega\beta_j \text{ ,} \\
    \frac{\partial\theta_{s,j}}{\partial t}-\kappa_j\frac{\partial^2\theta_{s,j}}{\partial z^2} & = 0
    \end{aligned}
\end{equation}
The continuity of temperature and thermal flux at $z=l$ requires
\begin{equation}\label{A4}
    \begin{aligned}
    \theta_{p,f}(z=l)+\theta_{1,f}\cosh(\gamma_f l)= \\
    \theta_{p,s}(z=l)+\theta_{1,s}\cosh(\gamma_s (h-l)) \text{ ,} \\ \\
    k_f[\theta^{'}_{p,f}(z=l)+\theta_{1,f}\gamma_{f} \sinh(\gamma_f l)]= \\
    k_s[\theta^{'}_{p,s}(z=l)-\theta_{1,s}\gamma_{s}\sinh(\gamma_s (h-l))]
    \end{aligned}
\end{equation}
$\theta^{'}_{p,f}$ and $\theta^{'}_{p,s}$ are zero for homogeneous film and substrate. Substitute in $\theta_{p,f}$ and $\theta_{p,s}$ and solve for $\theta_{1,f}$, we get
\begin{equation}\label{A5}
    \begin{aligned}
    \theta_{1,f} &= \frac{\Delta\beta}{\cosh(\gamma_{f}l)+R\sinh(\gamma_{f}l)\coth(q)} \text{ ,} \\
    \theta_{1,s} &=
    -\frac{\Delta\beta R}{\coth(\gamma_{f}l)\sinh(q)+R\cosh(q)}
    \end{aligned}
\end{equation}
where $q = \gamma_{s}(h-l)$, $\Delta\beta = \beta_f-\beta_s$ and $R=(k_{f}\gamma_{f})/(k_{s}\gamma_{s})=\sqrt{(k_{f}C_{f})/(k_{s}C_{s})}$.

The thermal field solutions can thus be expressed as a sum of the particular and the specific solutions and be written as
\begin{equation}\label{A6}
    \begin{aligned}
    \theta_{f} &= -\beta_{f}+\frac{\Delta\beta\times\cosh(\gamma_{f}z)}{\cosh(\gamma_{f}l)+R\sinh(\gamma_{f}l)\coth(q)} \text{ ,}\\
    \theta_{s} &= -\beta_{s}-\frac{\Delta\beta R\times\cosh[\gamma_s (h-z)]}{\coth(\gamma_{f}l)\sinh(q)+R\cosh(q)}
    \end{aligned}
\end{equation}

\section{Energy dissipation due to thermoelastic response}
\label{sec:appendix-energydissipation}
Substituting Eq.~\ref{eq20} and the expressions obtained previously in Section~\ref{appliedelasticfield} into Eq.~\ref{eq21}, we get
\begin{equation}\label{B1}
\begin{aligned}
    E_{diss,f} &= \frac{2\pi}{\omega}\int_{0}^{l}\omega\sum \operatorname{Im}[\sigma_{1,ii}\epsilon_{0,ii}+\sigma_{0,ii}\epsilon_{1,ii}]dz\\
    &=2\pi\sigma_{0}\int_{0}^{l}\operatorname{Im}[d_{0,f}]dz
\end{aligned}
\end{equation}
where $d_{0,f,\parallel}=B_{1}a_{0}+b_{0}A_{1}=-(2-2\nu_{f})\alpha_{f}\theta_{f}+(4-2\nu_{f})\alpha_{s}\theta_{s,l}$ and $d_{0,f,\perp}=B_{1}c_{0}+d_{0}A_{1}=\alpha_{f}\theta_{f}+2\nu_{f}(\alpha_{s}\theta_{s,l})/(1-\nu_{f})$.

In previous calculations in Section~\ref{formulationthermalfield}, we have shown that $\theta_{p,f}$ is a constant and a real number. Hence Equation~\ref{B1} can be simplified into
\begin{equation}\label{B2}
    \begin{aligned}
    E_{diss,f,\parallel} &= 2\pi\sigma_{0}\times \operatorname{Im}\left[(2\nu_{f}-2)\frac{\alpha_{f}\theta_{1,f}}{\gamma_{f}}\sinh(\gamma_f l) \right. \\
    &\quad +(4-2\nu_{f})\alpha_{s}\theta_{1,s}\cosh(q)\times l\Big. \Big] \text{ ,} \\
    E_{diss,f,\perp} &= 2\pi\sigma_{0}\times \operatorname{Im}\left[\frac{\alpha_{f}\theta_{1,f}}{\gamma_{f}}\sinh(\gamma_f l) \right. \\
    &\quad +\frac{2\nu_{f}}{1-\nu_{f}}\alpha_{s}\theta_{1,s}\cosh(q)\times l\Big. \Big]
    \end{aligned}
\end{equation}
where $q = \gamma_{s}(h-l)$.

A similar derivation for the substrate can be done by following all the procedures discussed above. The energy dissipated in the substrate can be determined and expressed in
\begin{equation}\label{B3}
    \begin{aligned}
    E_{diss,s} &= \frac{2\pi}{\omega}\int_{l}^{h}\omega\sum \operatorname{Im}[\sigma_{1,ii,s}\epsilon_{0,ii,s}+\sigma_{0,ii,s}\epsilon_{1,ii,s}]dz\\
    &=2\pi\sigma_{0}\int_{l}^{h}\operatorname{Im}[d_{0,s}]dz\\
    \end{aligned}
\end{equation}
where $d_{0,s,\parallel}=2\alpha_{s}\theta_{s}$ and $d_{0,s,\perp}=(1+\nu_{s})\alpha_{s}\theta_{s}/(1-\nu_{s})$. And
\begin{equation}\label{B4}
    \begin{aligned}
    E_{diss,s,\parallel} &=2\pi\sigma_{0}\times\operatorname{Im}\left[-2\frac{\alpha_{s}\theta_{1,s}}{\gamma_{s}}\sinh(q)\right]\\
    E_{diss,s,\perp} &=2\pi\sigma_{0}\times\operatorname{Im}\left[-\frac{1+\nu_{s}}{1-\nu_{s}}\frac{\alpha_{s}\theta_{1,s}}{\gamma_{s}}\sinh(q)\right]
    \end{aligned}
\end{equation}

In order to calculate the thermoelastic loss in the film and the substrate, the total energy dissipated and the total elastic energy stored have to be found. The total energy dissipated is simply the sum of $E_{diss,f}$ and $E_{diss,s}$ and is given by
\begin{equation}\label{B5}
    \begin{aligned}
        E_{diss,total,\parallel} &= E_{diss,f,\parallel}+E_{diss,s,\parallel}\\
        &= 2\pi\sigma_{0}\times \operatorname{Im}\left[(2\nu_{f}-2)\frac{\alpha_{f}\theta_{1,f}}{\gamma_{f}}\sinh(\gamma_f l) \right.\\
        &\quad +(4-2\nu_{f})\alpha_{s}\theta_{1,s}\cosh(q)\times l\\
        &\quad -2\frac{\alpha_{s}\theta_{1,s}}{\gamma_{s}}\sinh(q)\bigg. \bigg]
    \end{aligned}
\end{equation}
\begin{equation}\label{B6}
    \begin{aligned}
        E_{diss,total,\perp} &= E_{diss,f,\perp}+E_{diss,s,\perp}\\
        &= 2\pi\sigma_{0}\times \operatorname{Im}\left[\frac{\alpha_{f}\theta_{1,f}}{\gamma_{f}}\sinh(\gamma_f l) \right.\\
        &\quad +\frac{2\nu_{f}}{1-\nu_{f}}\alpha_{s}\theta_{1,s}\cosh(q)\times l\\
        &\quad -\frac{1+\nu_{s}}{1-\nu_{s}}\frac{\alpha_{s}\theta_{1,s}}{\gamma_{s}}\sinh(q)\bigg. \bigg]
    \end{aligned}
\end{equation}

 \bibliographystyle{elsarticle-num} 
 \bibliography{references}





% \end{thebibliography}
\end{document}